\documentclass[10pt,journal,compsoc]{IEEEtran}
\usepackage{listings}
\usepackage{url}

\ifCLASSOPTIONcompsoc
  \usepackage[nocompress]{cite}
\else
  \usepackage{cite}
\fi
\ifCLASSINFOpdf
  \usepackage[pdftex]{graphicx}
  \DeclareGraphicsExtensions{.pdf,.jpeg,.png}
\else
\fi
\begin{document}
%
\title{What Machine Learning can learn \\ from Software Modularity}
%
%
%
%

\author{Peter Kriens
        and Tim Verbelen}

\IEEEtitleabstractindextext{%
\begin{abstract}
In the last couple of years we have witnessed an enormous increase of machine learning (ML) applications. More and more program functions are no longer written in code, but learnt from a huge amount of data samples using an ML algorithm. However, what is often overlooked is the complexity of managing the resulting ML models as well as bringing these into real production systems. In software engineering, we have spent decades on developing tools and methodologies to create, manage and assemble complex software modules. We present an overview of current techniques to manage complex software, and how this applies to ML models.

\end{abstract}


}

\maketitle

\IEEEdisplaynontitleabstractindextext

%
\IEEEpeerreviewmaketitle

\IEEEraisesectionheading{\section{Introduction}\label{sec:introduction}}

%
%
%
%
\IEEEPARstart{I}n the past two decades we have seen an enormous proliferation of machine learning (ML) applications, mainly due to the success of deep learning algorithms in the fields of image processing, speech recognition and machine translation~\cite{Goodfellow2016}. Also in academia one witnesses an exponential growth of research papers on machine learning techniques. 

However, what is often overlooked is the complexity of managing a pipeline to learn these models as well as bringing these state of the art machine learning techniques into a real production system~\cite{Nascimento2019}. The highly experimental workflow of training an ML model, where a huge amount of hyperparameters can be tested and tweaked, leads to the widespread use of anti-patterns such as ``glue code'', ``(build) pipeline jungles'' and many dead experimental code paths~\cite{Sculley2015}. For example, Lin and Ryaboy~\cite{Lin2013} describe their work of building the Twitter big data mining infrastructure as mainly \textit{plumbing}. Although plumbing is usually not regarded as high tech, the consequences of bad plumbing can be equally unpleasant in real life as in software.

\section{Machine learning workflow}

\begin{figure}[b!]
   \centering
   \includegraphics[width=\linewidth]{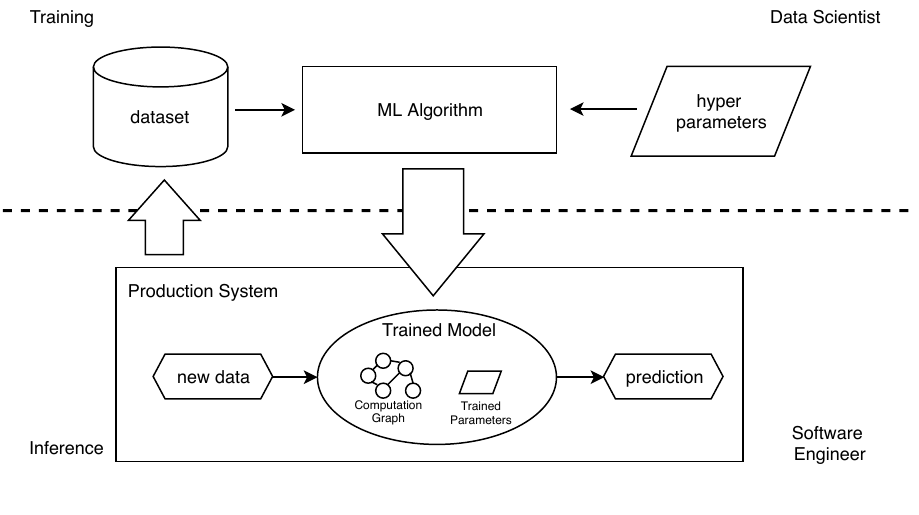}
   \caption{A typical machine learning workflow. During training a dataset and hyperparameters are fed into an ML algorithm which produces a trained model consisting of a computation graph and trained parameters. At inference time the trained model is ingressed with new data and the resulting predictions are used in the production system.}
   \label{fig:mlworkflow}
\end{figure}

In general one can identify two phases in developing and deploying a machine learning model, as shown in Figure~\ref{fig:mlworkflow}. It all starts with the availability of \textit{data}, which is often extracted from a production system. This data needs to be cleaned, transformed to a common format, and be made available as a \textit{dataset} to work with. This dataset can then be consumed by a machine learning algorithm that \textit{trains} an \textit{ML model} by adjusting its parameters to best fit the dataset. A typical neural network is comprised of millions of parameters that need to be optimized, and in addition also the ML algorithm itself is influenced by a number of \textit{hyperparameters}. Something as small as the random seed for the learning phase can have a significant impact on the training outcome~\cite{Henderson2018}. Deciding on these hyperparameters is usually the realm of the (well paid) data scientists that ``work their magic'' on the data using programming languages like Python and R.

The result, which we call the ML model, typically consists of on the one hand the values of the trained parameters, and on the other hand the computation graph that defines how these parameters are used in a number of operations for transforming the input. Once the ML model is trained, it can be used to issue predictions based on incoming new data, also called \textit{inference}. However, deploying an ML model involves \textit{industrializing} it. The model has to be integrated with the existing production system(s): it will depend on software modules to ingress the data, and other modules will depend on the ML model predictions for further processing. Also, it has to become robust, must be resilient to simple faults, and should also adhere to all service level agreements in place. Deploying ML models is typically the job of \textit{software engineers}.

Clearly, an ML model now starts to look like any other software module, and we should take care of the \textit{software engineering} aspects: it has dependencies that need to be managed, it needs to be maintained over time while it is continuously evolving, it has requirements on the runtime, etc. If we treat it as a software artifact then what lessons can we learn from software engineering?

\section{What makes a software module?}

Software engineering, even though being a relatively young engineering discipline, has matured in the past few decades. Today, applications are routinely built from hundreds, sometimes thousands, of open source and in-house libraries. For example, Maven Central is a Java repository that hosts millions of freely available libraries. Many of those libraries are of high quality and are widely used in critical enterprise applications. NPM is a free repository for Javascript/Nodejs libraries that has hundreds of thousands of packages. Python has PyPI, and almost any popular language provides a central repository with reusable libraries. The amount of reuse of software libraries today went way beyond what was deemed possible a mere 20 years ago.

This transition has not been painless and many lessons were learned along the way. The primary lesson software repositories have taught us is that the \textit{packaging} is quite crucial. All these repositories provide standardized formats and \textit{metadata}. Data that describes an artifact in machine readable form and provides information about its dependencies. 

In the following sections we discuss the most important aspects that made software modules so successful. In these sections we use the term \textit{module} for a  software project that evolves over time and a \textit{revision} a released \textit{version} of that module. A \textit{library} is a module intended to be reused by other modules.

\subsection{Semantic Versioning}

The basic function of a \textit{version} is to be able to discriminate between two \textit{revisions} of the same module. Initially a version was an opaque string that humans could use to see which revision was older. Since the version was usually the only piece of metadata that survived different stages in the development pipeline, developers tried to cram more and more information in the version string. 

One of the primary pieces of metadata that was put in the version was the backward compatibility status of a revision with respect to its predecessors. Was a new revision a simple drop in replacement or would it require changes in the code of its users? After many \textit{version wars} (developers are surprisingly picky about versions, we personally found out) over the syntax of the version format, the industry seemed to finally have settled on a standard promoted by Github called \textit{semantic versioning}. From the semver.org website~\cite{Semver}:

Given a version number MAJOR.MINOR.PATCH, increment the:

\begin{itemize}
\item MAJOR version when you make incompatible API changes,
\item MINOR version when you add functionality in a backwards-compatible manner, and
\item PATCH version when you make backwards-compatible bug fixes.
\end{itemize}

This means that if a library A is built against version $1.2.3$ of library B then library A can be used in the same runtime with B, as long as B has a version in the range of $[1.2.3,2.0.0)$. If A is ever in the same runtime with a B that has version $2.1.8$ then there is a potential problem. In that case, B is not guaranteed to work with A.

Since semantic versions can be interpreted by tools, it enables more verification in runtime to ensure that modules are actually compatible with each other. Clearly a major problem is that humans are notoriously bad in maintaining versions. Humans need tools to do this correctly once a system reaches a certain size.

\subsection{Dependencies}

\begin{figure}[h]
   \centering
   \includegraphics[width=2in]{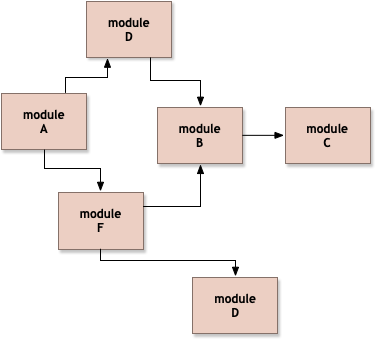}
   \caption{The transitive dependencies of module A are the direct dependencies D and F, as well as all recursive dependencies of those.}
   \label{fig:transitive}
\end{figure}

Few modules are an island, therefore, in the majority of cases modules depend on other libraries. These dependencies together create a graph. Virtually all repositories provide metadata that can refer to other libraries that must be included in the assembly. Since it is so easy to depend on other libraries the graph is often surprisingly large, especially in open source. A popular library in the Java world is Spring. Once you use one module, you use all its modules. This is a problem as large transitive depency graphs can pose constraints that are hard or sometimes impossible to solve. Figure~\ref{fig:transitive} shows how modules transitively depend on each other.

\subsection{Runtime Assembly}

At the end, all software artifacts need to be deployed together in a runtime. In the Unix world it became common to compile the modules on the target system. This allows the code to be optimized for the target but it creates a nightmare of complexity to synchronize the build environments of all machines that deploy the target. Few things are more frustrating as installing an enticing application and seeing a Python exception pointing out that the C compiler failed on one of the hundreds of C sources somewhere deep in a highly complex Makefile. Kudos to the developers that provide us with those applications but for large scale industrial applications it just does not work that well. In that world, binaries are required to keep the deployments manageable.

Reusing external libraries in binary form has the advantage that the (often significant) complexity of their build systems is not visible to the users of those libraries. Depending on binaries, developers can use a completely different toolsets to build their applications than the developers of the libraries they rely upon. This is a hard to overestimate reduction in complexity. Although there are some new challenges, \textit{well managed} binaries provide serious advantages in many stages of development pipeline. 

One of those new challenges is to \textit{assemble} the application out of all its parts so that it can run on its target environment. The \textit{assembly phase} collects all the libraries and transforms them into one or more artifacts that the target can execute. This can for example be an executable for Linux but also a deployment package for Kubernetes. When this assembled application consists of hundreds or thousands of libraries that each have their own web of dependencies then it is paramount that tools can verify that all modules are actually \textit{compatible} with all other modules. 

To be able to assemble applications reliably, it is absolutely necessary to define the structure and metadata that is used through the whole development pipeline. Software repositories like Maven and NPM have standardized the metadata for module naming and versioning.

\section{How about ML Models?}

Ultimately, ML models are software too, and it appears that, very similarly, we also have to manage their version, dependencies and runtime assembly. In contrary to software modules however, these concepts seem so far less established in the field of data science. 

Managing compatibility is a non trivial task since quite often the data scientists continue to tweak the ML model while it is being deployed. There is a very strong coupling between the trained ML model and the format of the input and the delivered output. The ML inference engine expects \textbf{exactly} the same data format as that what was used to train the ML model. Even the slightest change can wreak havoc. Although ML models can be surprisingly forgiving for fuzzy input, they are furies when it comes to the wrong data format. 

For example, if the data is trained on images with a resolution of 224x224 pixels then the ML model will most often only provide answers for that resolution. When the input image to the inference engine is 1280x960 then this has to be scaled or cropped to match the resolution at training time. The same is true for the output side. If the data scientists decided to change the classification then the module that takes the predictions must be adapted to accept those new classes.

The problem is even more apparent when resolution matches, but the semantic meaning does not. In contrast with almost any software module, ML models do not validate input nor output: any matching bit pattern is fair game. If you input an image encoded as floating point numbers naively rescaled between zero and one, while the ML model was trained on images normalized such that your dataset images have mean zero, then the model will give you an output, but you have no idea whether this is a valid output, most probably not. For example it has been shown that by adding a very small amount of carefully constructed noise you can get an image that looks exactly like a panda to a human, to be classified by a neural network as a ``gibbon''~\cite{Goodfellow2014}. Diagnosing these problems can be a nightmare since there are many reasons the output of an ML inference engine can resemble gibberish.

Hence it is clear that, in addition to source code dependencies, an ML model has a more subtle, yet more important dependency on data - the dataset used to train the model. This makes them increasingly harder to manage. Reproducible builds as we know them from software engineering become much more harder, if not impossible. You can easily check out a version of your source code from your code versioning system, not so from your billion records (and growing) dataset.

Also the phase of packaging and assembly of ML models is often overlooked. For example, trained neural networks are generally stored in a custom format depending on the framework used. In Pytorch, the recommended way is to store only the trained parameters~\cite{Pytorch}, which assumes the model code is carved in stone and will not change when the parameters are loaded in the future. TensorFlow stores both the computation graph and the trained parameters in a portable format, and also provides a serving mechanism to expose the ML model as a REST endpoint~\cite{Tensorflow}. TensorFlow serving also introduces a very limited concept of versioning by dumping each new model in a separate subdirectory on the filesystem with a monotonically increasing integer as name. The Open Neural Network eXchange (ONNX) format aims to provide a platform-agnostic format for storing both the computation graph and trained parameters~\cite{Onnx}. Although this is a step in the right direction - providing a generic packaging for a model - there is still a lack of declaring all necessary dependencies and metadata. 

\section{Requirements and Capabilities}

Trying to map ML models to the standard version and dependency models available in software, like for example Maven, left one of us not satisfied with the existing solutions. In another project he had used OSGi to develop applications and expected that some of the mechanisms developed in OSGi could be extremely useful. 

\subsection{OSGi}

OSGi is a specification for Java based applications that provides a very comprehensive life cycle model for reusable module based software engineering. It is very mature and there are many different implementations available. A well known product based on OSGi is the Eclipse IDE.

One of the unique aspects of OSGi is the treatment of dependencies. Instead of the common and widespread model where each \textit{module} depends on a set of other modules transitively, the OSGi chose to model dependencies on the concept of \textit{capabilities}~\cite{OSGi2012}.

In OSGi, modules never depend on other modules, they depend on \textit{APIs}. That is, the API became an independent reified entity provided by a module. Although separate API's have been best practice for quite some time, the idea to make the API a first class citizen in the dependency model is new as far as we know. The success of this \textit{API as a dependency anchor} made the OSGi adopt a model where modules have many different \textit{types} of dependencies. 

The OSGi therefore took the innovative step to abstract dependency \textit{types} into the \textit{Requirements and Capabilities specification}. This specification defines how generic capabilities and requirements are encoded and allows almost anything that a developer can make assumptions about to be specified as a requirement on capabilities so that it can be verified by tools. 

Capabilities and requirements are typed by a \textit{namespace}, which defines the meaning of a set of \textit{properties}. A namespace is similar to a type in a programming language. Properties are simple key/value pairs, where the values can only contain a very restricted set of types. The OSGi defines a number of namespaces for their own uses but any developer can develop their own namespaces, the model is fully extensible.
 
A \textit{capability} is a namespace and a set of properties. A \textit{requirement} is a namespace and a \textit{filter} expression that is asserted on the properties of the capabilities in the same namespace. The expressive filter language supports conjunction (and), disjunction (or), negation (not), and subexpressions to any reasonable depth. 

\begin{figure}[h]
   \centering
   \includegraphics[width=1.6in]{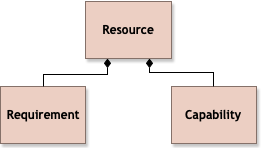}
   \caption{A Resource is defined by its Capabilities - what it offers - and its Requirements - what it needs to function.}
   \label{fig:resource}
\end{figure}

Finally, \textit{Resources} (Fig.~\ref{fig:resource}) are the modules that group a set of requirements and capabilities. Resources can be any binary file, they are not restricted to be OSGi modules.

Resources are assembled into a runtime using a \textit{Resolver}. A resolver takes a set of \textit{initial requirements} and repositories with Resources. Since each resource is described by capabilities and requirements, the Resolver can calculate a closure of Resources that have no unsatisfied requirements. This closure is an assembly that can be safely deployed.

For example, a Java class might have an import statement to depend on a package ``com.example.foo''.
Tooling therefore adds a \textit{package} requirement to its module. This would look something like the following in the metadata:

\begin{lstlisting}
osgi.wiring.package;
  filter:='(osgi.wiring.package=com.example.foo)'
\end{lstlisting}

The resolver will then select a module that has the following capability:

\begin{lstlisting}
osgi.wiring.package; 
  osgi.wiring.package='com.example.foo'
\end{lstlisting}

For most developers in the trenches this sounds unrealistic since they are used to the myriad of ways that things fail. It is hard to imagine for them that the enormous complexity they are faced with daily could be helped by such a relatively simple model. It is true that reaching ``cruise level'' with the Requirements Capability model is an uphill battle. Getting started with it shows the enormous amount of unmanaged assumptions that are made in existing software. Assumptions that are so often the cause of bugs. Although there are now many tools for managing and generating the metadata, fixing the until then unreported bugs is a major task. However, when civil engineers build bridges, they also do not complain that all structural parts need to be calculated for many parameters. Software has always been a very non-engineering task that has been relying more on human ingenuity than proper engineering.

\subsection{Potential usage in ML models}

As a prototype, we also applied this Requirements Capability model to neural network models ~\cite{Dianne}. The resulting artifact, in casu a `.jar` file, contains not only the definition of the computation graph and the trained parameters, but also several capabilities of the model. For example, what does the model accept as input? I.e. images of a certain width x height resolution. What does the model yield as output? A set of labels of a classification problem. Another capability defines the dataset the model was trained on. 

An example capability could look like this:

\begin{lstlisting}
 ml.model;
   input=image;
   input.height:Long=28;
   input.width:Long=28;
   output.type=digit;
   output.size:Long=10;
   dataset=MNIST;
   version:String=1.0.0
\end{lstlisting}

The previous capability defines an ML model that is trained on the MNIST dataset, which consists out of 28x28 images of digits between 0 and 9. 

The model also defines a set of requirements: which packages are required that provide implementations of the necessary operations for executing the computation graph. One might also add hardware requirements in terms of GPU availability or memory.

The real benefit becomes apparent when writing a consumer. On the consumer side, requirements can be declared, for example on the resolution of images that need to be processed, and which types need to be classified. Using the OSGi Resolver, one can then automatically assemble a set of artifacts that \textit{resolve} a transitive closure of the consumer's requirements, containing a suitable ML model, together with all necessary implementations to actually execute this on the hardware at hand. It is hard to overestimate the value of moving these type of detailed dependency handling away from humans.

For example a consumer might define the following requirement:

\begin{lstlisting}
   ml.model;
      filter:='(\& 
         (input=image)
         (input.width>=28)
         (input.height>=28)
         (output.type=digit)
         (|(dataset=MNIST)(dataset=SVHN)))'
\end{lstlisting}

This requirement states that it wants to process images and classify these as digits, but it takes models that are trained on either the MNIST or SVHN dataset.

\subsection{Future outlook}

Despite the apparent need of developing better software engineering practices in light of the machine learning evolution, those challenges still remain~\cite{Khomh2018}. Recent surveys~\cite{Lwakatare2019,Zhang2019, Giray2021} still list many open issues regarding ML model deployment, integration and maintenance. Due to their dependency on the data they are trained with, ML models may exhibit non-monotonic error behavior and are much more difficult to test and integrate~\cite{Amershi2019}. Meanwhile, other scientists also stress the importance of more carefully curating metadata for datasets~\cite{Gebru2020} and many other (hyper)parameters involved in ML model training~\cite{Schelter2017}. We believe further formalizing such metadata as requirements and capabilities, which enables the automatic resolving of compatible components, will be indispensable as system complexity further increases.

\section{Conclusion}

In the seminal book the Mythical Man Month~\cite{Brooks1995}, Brooks describes how a two men garage team can do in days what professional teams do in months. He explains the discrepancy that a \textit{product} is much more than the software alone. All developers know how easy it is to get a program working and how much drudgery work is involved in industrializing that program into a product. Experience has taught us the importance of carefully modularizing, packaging and assembling software, e.g. by adding version and dependency metadata. 

The OSGi Requirements/Capability model is the first model we know that tries to provide generic metadata so that all stages of development can verify the assumptions of earlier stages with tooling, preventing a surprisingly large number of errors in the later stages of development and automating traditionally cumbersome tasks. 

With the proliferation of machine learning, increasingly many program functions are no longer captured in code, but learnt from a dataset and an ML algorithm. However, in the total workflow of creating such an ML model, the deployment phase often comes as an afterthought. Still, the resulting ML models experience the same deployment model as many other software modules. In addition, their unique characteristics can make them even more complex to deploy in production. By explicitly declaring the capabilities and requirements, and automatically resolving them at development time, many downstream problems with the models can be prevented before they become visible to the customer. This urges for a new packaging model for ML models that captures all necessary metadata to automate the assembly process.

\ifCLASSOPTIONcaptionsoff
  \newpage
\fi



%
\newpage
\bibliographystyle{IEEEtran}
\bibliography{references}

\begin{thebibliography}{10}
\providecommand{\url}[1]{#1}
\csname url@samestyle\endcsname
\providecommand{\newblock}{\relax}
\providecommand{\bibinfo}[2]{#2}
\providecommand{\BIBentrySTDinterwordspacing}{\spaceskip=0pt\relax}
\providecommand{\BIBentryALTinterwordstretchfactor}{4}
\providecommand{\BIBentryALTinterwordspacing}{\spaceskip=\fontdimen2\font plus
\BIBentryALTinterwordstretchfactor\fontdimen3\font minus
  \fontdimen4\font\relax}
\providecommand{\BIBforeignlanguage}[2]{{%
\expandafter\ifx\csname l@#1\endcsname\relax
\typeout{** WARNING: IEEEtran.bst: No hyphenation pattern has been}%
\typeout{** loaded for the language `#1'. Using the pattern for}%
\typeout{** the default language instead.}%
\else
\language=\csname l@#1\endcsname
\fi
#2}}
\providecommand{\BIBdecl}{\relax}
\BIBdecl

\bibitem{Goodfellow2016}
I.~Goodfellow, Y.~Bengio, and A.~Courville, \emph{Scaling Learning Algorithms
  Towards {AI}}.\hskip 1em plus 0.5em minus 0.4em\relax MIT Press, 2016.

\bibitem{Nascimento2019}
\BIBentryALTinterwordspacing
E.~Nascimento, I.~Ahmed, E.~Oliveira, M.~Palheta, I.~Steinmacher, and T.~Conte,
  ``Understanding development process of machine learning systems: Challenges
  and solutions,'' in \emph{2019 ACM/IEEE International Symposium on Empirical
  Software Engineering and Measurement (ESEM)}.\hskip 1em plus 0.5em minus
  0.4em\relax Los Alamitos, CA, USA: IEEE Computer Society, sep 2019, pp. 1--6.
  [Online]. Available:
  \url{https://doi.ieeecomputersociety.org/10.1109/ESEM.2019.8870157}
\BIBentrySTDinterwordspacing

\bibitem{Sculley2015}
D.~Sculley, G.~Holt, D.~Golovin, E.~Davydov, T.~Phillips, D.~Ebner,
  V.~Chaudhary, M.~Young, J.-F. Crespo, and D.~Dennison, ``Hidden technical
  debt in machine learning systems,'' in \emph{Proceedings of the 28th
  International Conference on Neural Information Processing Systems}.\hskip 1em
  plus 0.5em minus 0.4em\relax MIT Press, 2015, pp. 2503--2511.

\bibitem{Lin2013}
\BIBentryALTinterwordspacing
J.~Lin and D.~Ryaboy, ``Scaling big data mining infrastructure: The twitter
  experience,'' \emph{SIGKDD Explor. Newsl.}, vol.~14, no.~2, pp. 6--19, Apr.
  2013. [Online]. Available: \url{http://doi.acm.org/10.1145/2481244.2481247}
\BIBentrySTDinterwordspacing

\bibitem{Henderson2018}
P.~Henderson, R.~Islam, P.~Bachman, J.~Pineau, D.~Precup, and D.~Meger, ``Deep
  reinforcement learning that matters,'' in \emph{AAAI}, 2018.

\bibitem{Semver}
T.~Preston-Werner, ``Semantic versioning 2.0.0,'' \url{https://semver.org/},
  2013.

\bibitem{Goodfellow2014}
I.~J. Goodfellow, J.~Shlens, and C.~Szegedy, ``Explaining and harnessing
  adversarial examples,'' \emph{arXiv preprint arXiv:1412.6572}, 2014.

\bibitem{Pytorch}
PyTorch, ``Saving and loading models in pytorch,''
  \url{https://pytorch.org/tutorials/beginner/saving_loading_models.html},
  2017.

\bibitem{Tensorflow}
{Google Inc.}, ``Serving a tensorflow model,''
  \url{https://www.tensorflow.org/serving/serving_basic}, 2018.

\bibitem{Onnx}
{Facebook Inc.}, ``The open neural network exchange format (onnx),''
  \url{https://onnx.ai/}, 2017.

\bibitem{OSGi2012}
{The {OSGi} Alliance}, ``{OSGi} service platform core specification, release
  5,'' \url{http://www.osgi.org/Specifications}, 2012.

\bibitem{Dianne}
T.~Verbelen, ``An osgi-powered neural network example,''
  \url{https://github.com/ibcn-cloudlet/dianne-examples/tree/master/dianne.examples.onnx},
  2018.

\bibitem{Khomh2018}
F.~Khomh, B.~Adams, J.~Cheng, M.~Fokaefs, and G.~Antoniol, ``Software
  engineering for machine-learning applications: The road ahead,'' \emph{IEEE
  Software}, vol.~35, no.~5, pp. 81--84, 2018.

\bibitem{Lwakatare2019}
L.~E. Lwakatare, A.~Raj, J.~Bosch, H.~H. Olsson, and I.~Crnkovic, ``A taxonomy
  of software engineering challenges for machine learning systems: An empirical
  investigation,'' in \emph{Agile Processes in Software Engineering and Extreme
  Programming}, P.~Kruchten, S.~Fraser, and F.~Coallier, Eds.\hskip 1em plus
  0.5em minus 0.4em\relax Cham: Springer International Publishing, 2019, pp.
  227--243.

\bibitem{Zhang2019}
T.~Zhang, C.~Gao, L.~Ma, M.~Lyu, and M.~Kim, ``An empirical study of common
  challenges in developing deep learning applications,'' in \emph{2019 IEEE
  30th International Symposium on Software Reliability Engineering (ISSRE)},
  2019, pp. 104--115.

\bibitem{Giray2021}
\BIBentryALTinterwordspacing
G.~Giray, ``A software engineering perspective on engineering machine learning
  systems: State of the art and challenges,'' \emph{Journal of Systems and
  Software}, vol. 180, p. 111031, 2021. [Online]. Available:
  \url{https://www.sciencedirect.com/science/article/pii/S016412122100128X}
\BIBentrySTDinterwordspacing

\bibitem{Amershi2019}
S.~Amershi, A.~Begel, C.~Bird, R.~DeLine, H.~Gall, E.~Kamar, N.~Nagappan,
  B.~Nushi, and T.~Zimmermann, ``Software engineering for machine learning: A
  case study,'' in \emph{2019 IEEE/ACM 41st International Conference on
  Software Engineering: Software Engineering in Practice (ICSE-SEIP)}, 2019,
  pp. 291--300.

\bibitem{Gebru2020}
T.~Gebru, J.~Morgenstern, B.~Vecchione, J.~W. Vaughan, H.~Wallach, H.~D.~I.
  au2, and K.~Crawford, ``Datasheets for datasets,'' 2020.

\bibitem{Schelter2017}
S.~Schelter, J.-H. Boese, J.~Kirschnick, T.~Klein, and S.~Seufert,
  ``Automatically tracking metadata and provenance of machine learning
  experiments,'' in \emph{Machine Learning Systems Workshop at NIPS}, 2017, pp.
  27--29.

\bibitem{Brooks1995}
F.~P. Brooks, Jr., \emph{The Mythical Man-month (Anniversary Ed.)}.\hskip 1em
  plus 0.5em minus 0.4em\relax Addison-Wesley Longman Publishing Co., Inc.,
  1995.

\end{thebibliography}

%
\begin{IEEEbiography}[{\includegraphics[width=1in,height=1.25in,clip,keepaspectratio]{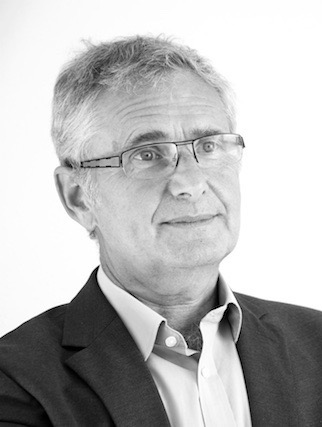}}]{Peter Kriens}


Peter Kriens was one of the key people developing OSGi since 1998. He was the editor of most specifications and was very instrumental in the development of OSGi as technical director. In 2015 he left the OSGi to successfully help companies to leverage OSGi to the max by providing mentoring and proof of concepts. he currently lives in the south of France.
\end{IEEEbiography}

\newpage

\begin{IEEEbiography}[{\includegraphics[width=1in,height=1.25in,clip,keepaspectratio]{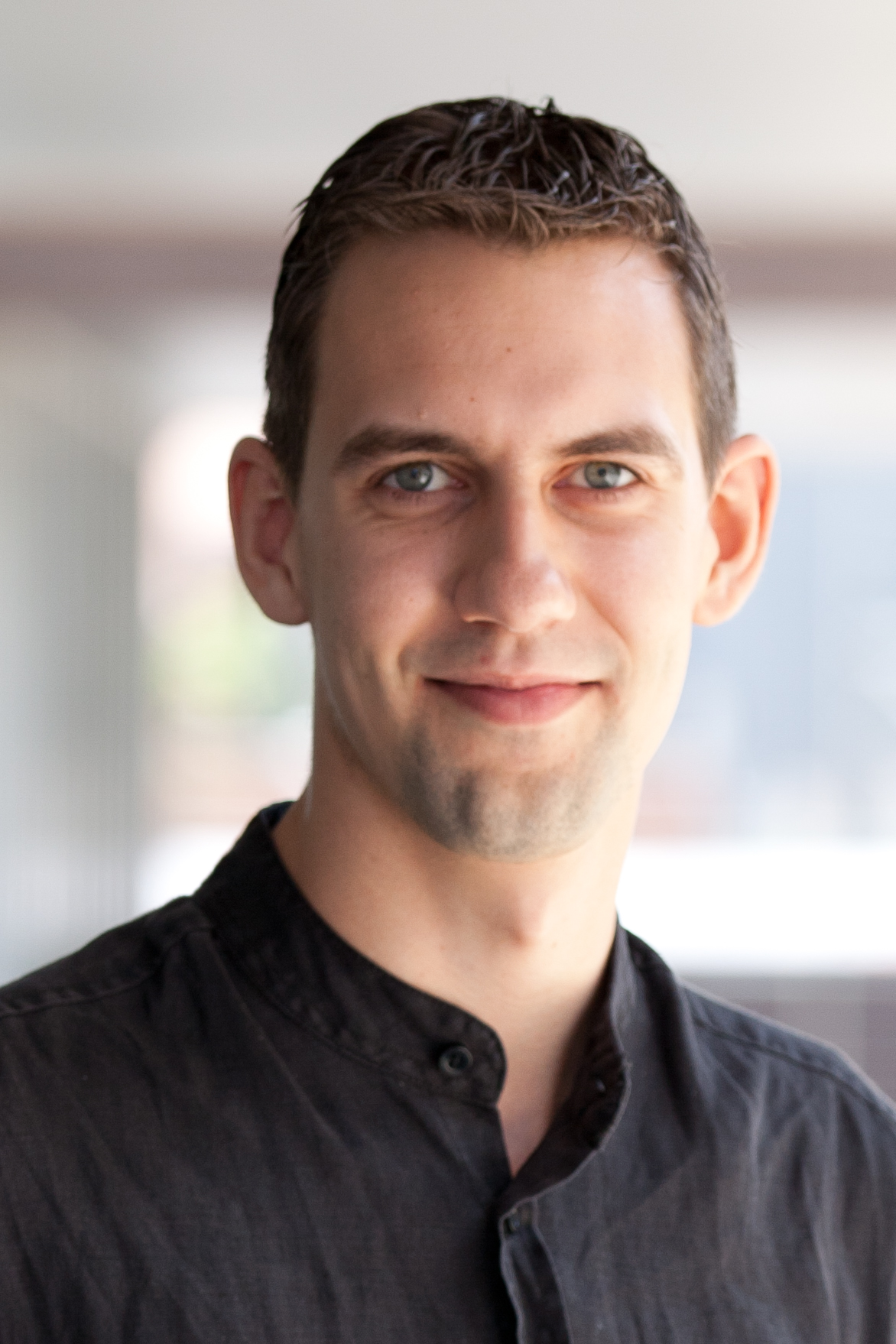}}]{Tim Verbelen}

Tim Verbelen received the M.Sc. and Ph.D. degrees in computer science engineering from Ghent University, Ghent, Belgium, in 2009 and 2013, respectively. Since then, he has been a Senior Researcher with Ghent University and IMEC, Leuven, Belgium. His main research interests include perception and control for autonomous systems using deep learning techniques, inspired by cognitive neuroscience theories such as active inference. Tim Verbelen was also an invited contributer of the OSGi Alliance from 2014 to 2020, and is Eclipse committer of the Concierge OSGi runtime project.
\end{IEEEbiography}






\end{document}